\documentclass[twocolumn]{aastex61}
\usepackage{epsfig}

\begin{document}
\title{The Maximum Isotropic Equivalent Energy Of Gamma Ray Bursts}

\author{Shlomo Dado}
\affiliation{Physics Department, Technion, Haifa 32000, Israel}
\author{Arnon Dar}
\affiliation{Physics Department, Technion, Haifa 32000, Israel}

\begin{abstract}
The canonball model, which unifies cosmic ray bursts (CRBs) and gamma ray 
bursts (GRBs), is used to predict the maximum isotropic equivalent gamma 
ray energy release in a GRB. The predicted maximum is based on the 
observed knee around 1 TeV in the energy spectrum of Galactic cosmic ray 
electrons, and on the Amati correlation in GRBs. Both were predicted by 
the cannonball model of CRBs and GRBs before their empirical discoveries. 
The predicted maximum agrees well with that concluded from uptodate GRB 
observations.

\end{abstract}

\section{Introduction}
Gamma ray bursts (GRBs) are the most luminous sources of electromagnetic 
radiation in the observable universe (Fishman \& Meegan 1995).  They were 
first detected on July 2, 1967 by the USA Vela spy satellites, which were 
launched to detect possible USSR tests of nuclear weapons above the 
atmosphere, in violation of the USA-USSR Nuclear Test Ban Treaty signed in 
1963. Their discovery was first published in 1973 after 15 such events 
were detected (Klebesadel et al. 1973), which have ruled out man-made 
origin and indicated that they were outside the solar system.

Until 1991, it was widely believed that the observable GRBs are located in 
our Galaxy. But, shortly after its launch in 1991, the Compton Gamma-Ray 
Burst Observatory (CGRO) provided compelling evidence that GRBs are 
extragalactic and their locations extend up to very large cosmological 
distances (Meegan, et al. 1992). Such cosmological distances and the 
prevailing assumption that the emitted radiation in GRBs is isotropic, 
implied that GRBs are the most energetic and luminous events in the 
universe since the big bang (Fishman \& Meegan 1995). Indeed, the 
discovery with the Italian-Dutch satellite BeppoSAX that GRBs have a 
longer-lived X-ray afterglow (Costa et al. 1997)  provided accurate enough 
sky localizations of GRBs, and led to the discovery of their optical 
afterglow (van Paradijs et al. 1997), their host galaxies and their 
redshifts, which confirm their enormous luminosities and isotropic 
equivalent energies, as implied by the observations (Meegan et al. 1992) 
with the Compton Gamma Ray Observatory (CGRO).

By now, the redshifts of more than 500 GRBs, out of nearly 2000 GRBs, 
which were located by the Compton, Konus/Wind, BeppoSAX, HETE2, INTEGRAL, 
Swift, AGILE, Fermi, CALET and AstroSat space based telescopes, have been 
measured with ground based telescopes and the Hubble space based 
telescope.  The distribution of the isotropic equivalent energy release, 
$E_{iso}$, of these GRBs show a strong cutoff beyond $\sim 1-3\times 
10^{54}$ erg (Atteia et al. 2017) with a largest observed value 
$E_{iso}\!=\! 3.7\times 10^{54}$ erg (Atteia et al. 2022), which was 
measured by Fermi/GBM (Lesage et al. 2022) in GRB 220101A at redshift 
$z\!=\!4.618$ (Fu et al. 2022, Fynbo et al. 2022). So far no GRB model has 
predicted nor explained the origin of the observed sharp cutoff/maximal 
value of $E_{iso}$ of GRBs.

In this letter we use two unique properties of cosmic-ray bursts (CRBs) 
and GRBs, which were {\bf predicted} by the cannonbal model that unifies 
CRBs and GRBs (Shaviv \& Dar 1995, Dar 1998, Dar \& Plaga 1999, Dar \& De 
R\'ujula 2000,2004,2008, Dado, Dar \& De R\'ujula 2022 and references 
therein), and have been confirmed by observations, to predict a  maximal 
value $\approx\! 3.8\times 10^{54}$ erg of $E_{iso}$ in GRBs. These two 
properties are the so called Amati correlation in GRBs (Amati et al. 2002, 
2006, 2009, 2019) and the knee around 1 TeV in the energy spectrum of 
cosmic ray electrons (Dado \& Dar 2015, De R\'ujula 2019), which was first 
indicated by the combined observations of the AMS-02 collaboration 
(Aguilar et al. 2014) and the H.E.S.S collaboration (Aharonian et al. 
2008,2009), and confirmed in more recent observations by the DAMPE 
collaboration (Ambrosi et al. 2017), and the CALET collaboration 
(Adriani et al. 2018).

\section{The Cannonball Model Of GRBs And CRBs}
In the cannonbal (CB) model of GRBs and CRBs bipolar jets of highly 
relativistic plasmoids (CBs) with an initial Lorentz factor $\gamma(0)\sim 
10^3$ are assumed to be launched by matter falling back onto a newly born 
compact stellar objects (a neutron star, a quark star or a black hole) in 
stripped envelope supernova explosions of type Ic (SNeIc) 
and in "failed supernovae" - direct collapse of a massive star to 
a black hole without a supernova (MacFadyen \& Woosley 1999). 

The electrons within CBs with a highly 
relativistic bulk motion produce prompt gamma-ray pulses by inverse 
Compton scattering (ICS) of photons of the light halo (glory) surrounding 
the progenitor star. Such a glory is produced by scattered light from 
pre-collapse ejecta, or from a companion star, or from an accretion disk 
formed around the compact stellar object. This CB model of GRBs has been 
extremly successful in {\bf predicting} the main observed properties of 
GRBs (e.g. Dado, Dar \& De R\'ujula 2022 and references therein).

CRBs are produced by the highly relativistic jets of CBs by scattering the 
particles on their path in the interstellar medium (ISM) to cosmic-ray 
energies (Dar \& plaga 1999, Dar\& De R\'ujula 2008). The highest energy 
that particles of a mass $m_i$ at rest in the ISM 
can be scattered to  by a CB with a Lorentz factor $\gamma(0)\gg 1$ is 
$\approx\! 2mc^2 [\gamma(0)]^2$. Further 
increase in their energy can  take place in the ISM if they happen to be
scattered by other CBs/fast moving matter in the ISM. Such 
secondary encounters  can raise their energy 
beyond the above limit, and turn it into a  CR knee in their energy 
spectrum, around an energy
\begin{equation} 
E_{knee}\approx 2m_ic^2[\gamma(0)]^2\,.              
\end{equation}
The knees in the energy spectrum of cosmic-ray 
nuclei of charge Ze  and mass  $\approx A\,m_p$
seem to satisfy 
\begin{equation} 
E_{knee}(A)\!\approx\! A\,E_{knee}(p),              
\end{equation}
where $E_{knee}(p)\approx 2$ PeV.
So far  measurments of the energy spectrum of cosmic ray nuclei
above the atmosphere at PeV energies  were not accurate enough to 
indicate whether the knee energy in their energy spectra  
is  proprtional to their rigidity, i.e. $E_{knee}(A)\!\approx\! 
Z\,E_{knee}(p)$  
as widely believed, or to their 
mass, as expected in the CB model (Dar \& De R\'ujula 2008). By now, 
this controversy seems to have been settled dramatically in favor of the 
CB model by the discovery of a knee 
in the energy spectrum of cosmic-ray electrons (CRe) around 
\begin{equation} 
E_{knee}(e)\!\approx\! (m_e/m_p)E_{knee}(p)\!\approx\! 1 {\rm TeV}.
\end{equation}
Evidence for a CRe knee around 1 TeV, was first suggested (Dado \& Dar 
2015) by combining the observations of CRe above TeV by the H.E.S.S 
collaboration (Aharonian et al. 2008, 2009) and by the observations of CRe 
below TeV by the AMS-02 collaboration (Aguilar et al. 2014),
as shown in Fig 1. It was later confirmed by the observations of DAMPE 
(Ambrosi et al. 2018) plus Fermi-LAT (Abdollahi et al. 2017) and 
by CALET (Adriani et al. 2018) plus AMS-02 (Aguilar al. 2014),
as shown in Figures 2 and 3, despite systematic differences in their 
spectra due to unknown origins. 

\section{The Maximum Isotropic Energy Of GRBs:}
In the CB model, (see e.g., Dado, Dar \& De R\'ujula 2022 
for a recent review) fall back material in SN explosions of 
type Ic (SNeIc) on the newly born compact object results in the 
ejection of a bipolar jet of highly relativistic plasmoids (CB) of 
ordinary matter with a large initial bulk motion Lorentz factor 
$\gamma(0)\sim O(10^3)$. 
Inverse Compton scattering (ICS) in the Thomson regime of an isotropic 
distribution of glory photons in the SNIc rest frame,  
with a typical peak energy $\epsilon_p\!\approx\!1$ eV, 
by the electrons 
in a jet of CBs with a typical initial Lorentz factor $\gamma(0)\sim 
O(10^3)$ in the SNIc rest frame at redshift $z$, yields a GRB photon 
distribution with a peak energy  $E_p$ in the observer frame that
satisfies,  
\begin{equation}
(1+z)\,E_p\!\approx\! \gamma(0)\delta(0)\epsilon_p\,,
\end{equation}
where $\delta(0)\!=\!1/[\gamma(0)(1\!-\!\beta cos\theta)]$ is the 
Doppler factor of the GRB viewed from an angle $\theta$ relative to the CB 
direction of motion. 
The isotropic equivalent  GRB energy in the SNIc rest frame 
satisfies  
\begin{equation}
E_{iso}\!\propto\!\gamma(0)\,[\delta(0)]^3\, \epsilon_p. 
\end{equation}
ICS of an isotropic photon distribution in the SN rest 
frame at redshift $z$, produces a GRB, which is beamed into an angular 
distribution $(dn_\gamma/d\Omega)\!\approx\!(n_\gamma/4\pi)\delta^2$. 
The mean scattereing angle of photons undergoing Compton 
scattering  is $\pi/2$, in the CB rest frame
or $\theta\!=\! 1/\gamma(0)$, in the observer frame.
It yields $\delta(0)\!\approx \!\gamma(0)$, and the Amati 
correlation   
\begin{equation}
(1\!+\!z)E_p\!\propto\! [E_{iso}]^{1/2}, 
\end{equation}
which follows from Eqs. (4) and (5). This CB model correlation 
is in excellent agreement with the latest best fit 
Amati correlation (Amati et al. 2019),   
\begin{equation} 
[(1\!+\!z)E_p/100{\rm keV})\!\approx \!115[E_{iso}/10^{52}{\rm 
erg}]^{0.50\!\pm\!0.02}, 
\end{equation}
which was discovered empirically, two decades ago, tested continuously 
and confirmed repeatedly with new observational data on 
GRBs (e.g., Amati et al. 2002,2006,2009,2019).

Single scattering of interstellar ionized 
particles (atomic nuclei of mass $m_i\!=\!m_A$ and electrons of mass
$m_e$, respectivly) on their path creates a highly 
relativistic beam of cosmic ray particles with maximum energies 
$E_{max}\!\approx 2m_i[\gamma(0)]^2$. In the CB model
these maximum energies of ISM particles  
acquired in a single scattering, are the knee energies in the  
energy spectra of cosmic ray nuclei and electrons.
The first indication of a knee around 1 TeV in the energy 
spectrum of cosmic ray electrons plus positrons (CRe) was obtained by
combining the CRe observations of the H.E.S.S collaboration
(Aharonian et al. 2008,2009) and of the AMS collaboration
(Aguilar et al. 2014) shown in Figure 1, although  the H.E.S.S 
results were qualified by sizeable sytematic uncertainties. 
\begin{figure}[]
\centering
\epsfig{file=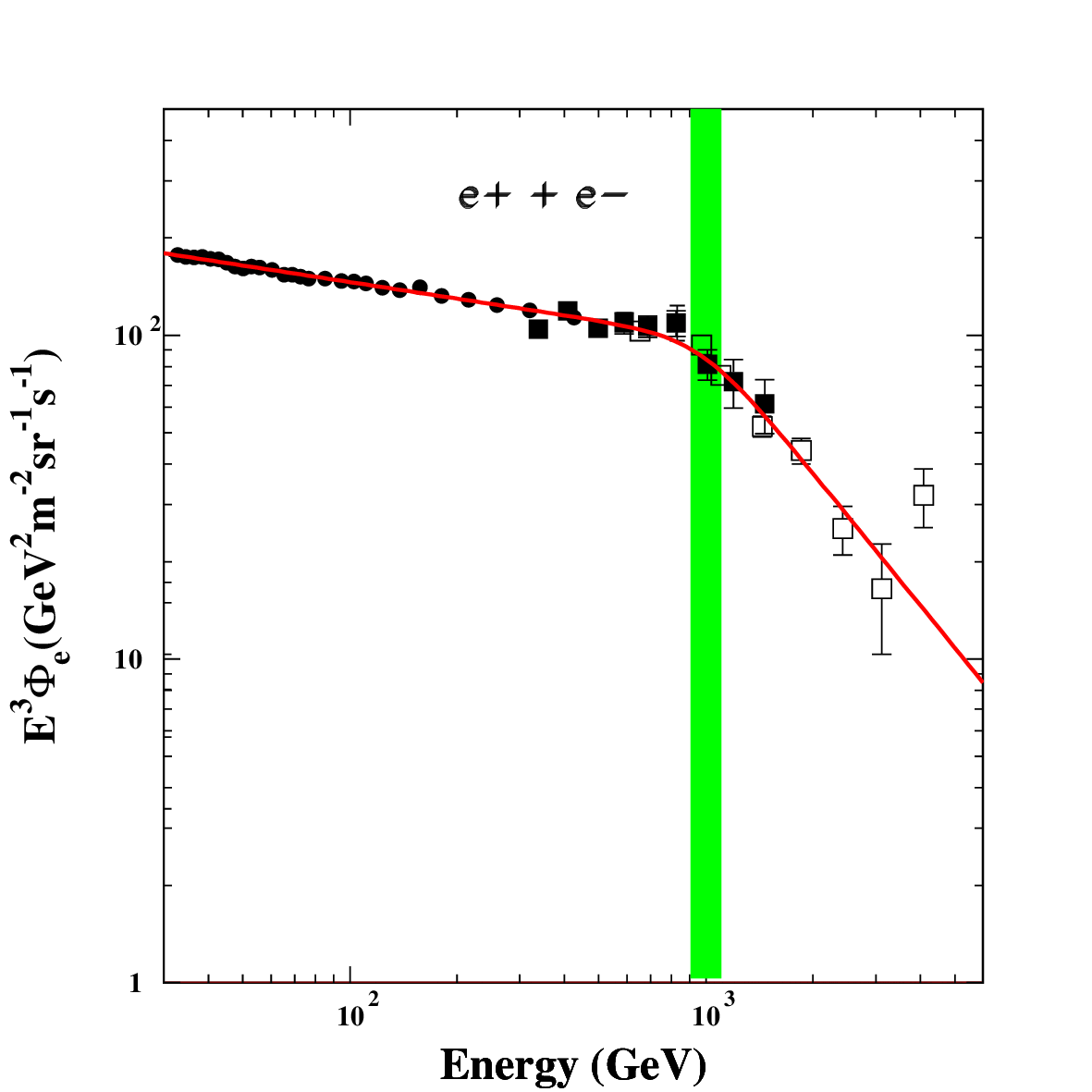,width=8.6cm,height=8.6cm}
\caption{A cutoff power-law fit to the combined CRe flux
measured near Earth with AMS (full circles: Aguilar et al. 2014)
and with H.E.S.S (squares: Aharonian et al. 2008,2009) 
The normalization of the H.E.S.S data was adjusted within their
estimated systematic error to match the more precise AMS-02 data 
below TeV (Aguilar et al. 2014)}
\end{figure}\\

The presence of a CRe knee around 1 TeV (Dado \& Dar  2015) in the energy 
spectrum of Galactic cosmic ray electrons plus positrons (CRe) was 
recently confirmed in extended CRe observations with DAMPE, the Dark 
Matter Particle Explorer (Chang et al, 2017)  and independently with 
CALET, the Calorimetric Electron Telescope (Adriani et al. 2018), shown in 
Figures 2 and 3, respectively.
\begin{figure}[]
\centering
\epsfig{file=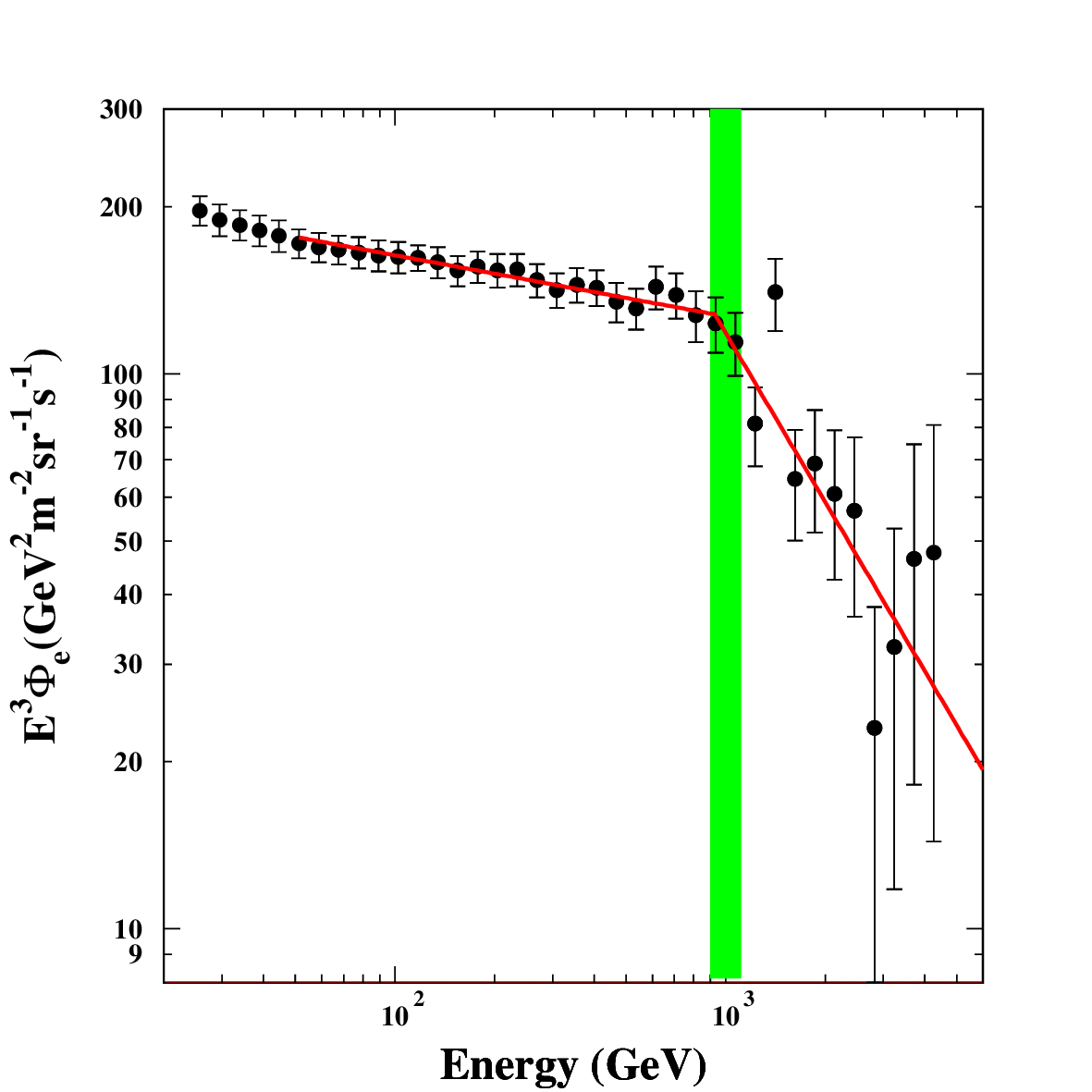,width=8.6cm,height=8.6cm}
\caption{: A broken power-law fit to the CRe spectrum (multiplied by 
$E^3$) measured by DAMPE (Chang et al. 2017) between 50 GeV - 5 TeV. 
A CRe knee is indicated by the wide band around 1 TeV.}
\end{figure}\\
\begin{figure}[]
\centering
\epsfig{file=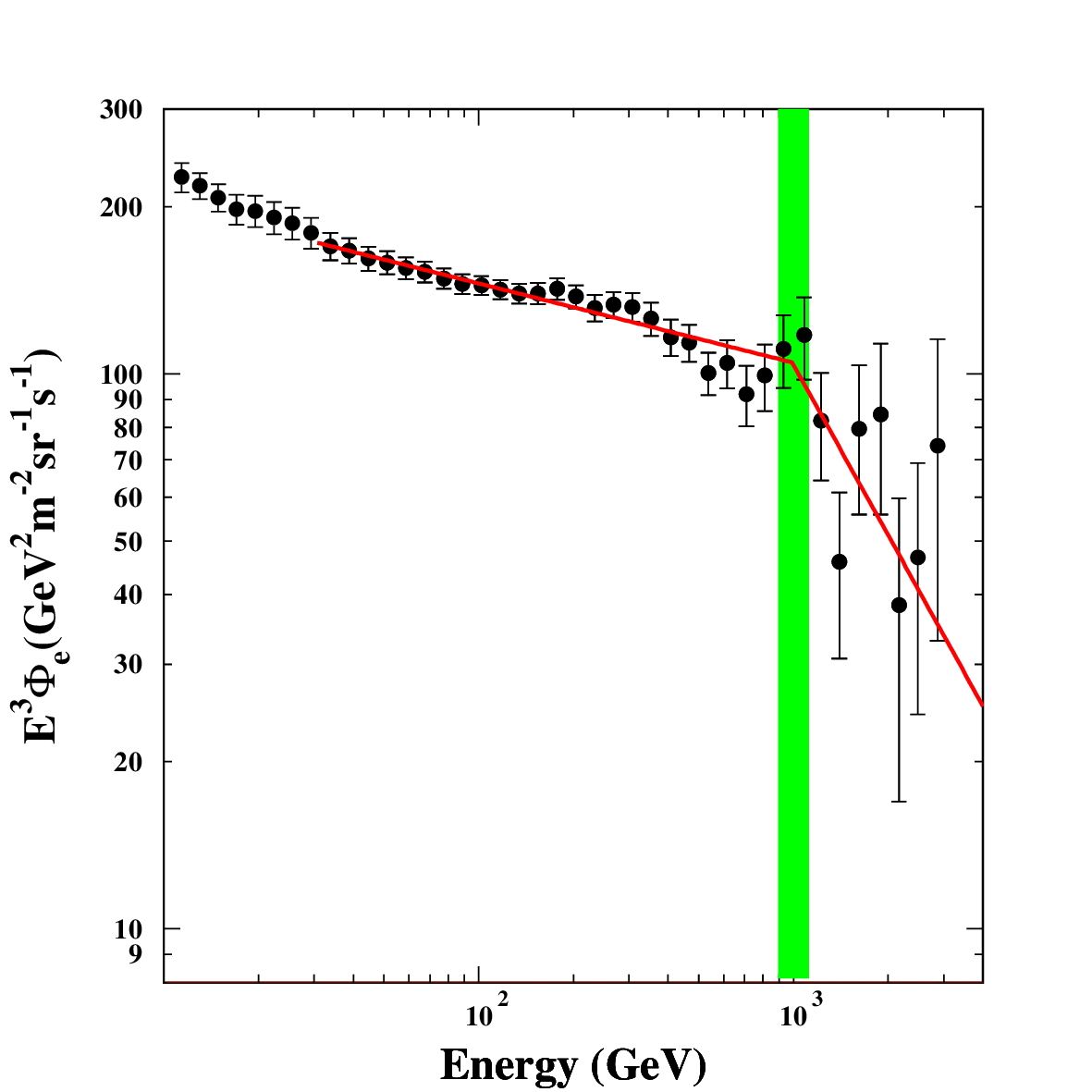,width=8.6cm,height=8.6cm}
\caption{A broken power law fit to the CRe
spectrum (multiplied by $E^3$) measured with the Calorimetric Electron 
Telescope (CALET) on the Intenational Space Station,  from   
11 GeV to 4.8 TeV.}
\end{figure}

A CRe knee around $\approx 1$ TeV  implies a maximum initial 
Lorentz factor $\gamma(0)\approx 1500$ of CBs. 
According to Eq.(4), ICS of glory 
photons of typical peak energy $\epsilon_p\!\approx\!1$ eV 
by inert electrons in CBs with $\gamma(0)\approx 1500$ yields  
\begin{equation} 
(1\!+\!z)E_p\!\approx\! 2.25\, MeV. 
\end{equation} 
This value of $(1\!+\!z)E_p$ and the best fit Amati correlation 
as given by Eq.(4) yield  
max $E_{iso}\!\approx\! 3.80\times 10^{54}$  erg. 
Strictly, this value corresponds to GRBs produced by ICS of 
glory photopns with a peak energy $\approx\!1$ eV by CBs 
moving at an angle $\theta\!\approx\!1/\gamma(0)$
relative to the line of sight to the GRB.
Taking into account the spreads in viewing angle and  
peak energy of glory photons, this value is actually the 
value beyond  which the observed distribution of $E_{iso}$ 
of GRBs is predicted to have a strong cutoff. This 
strong cutoff is evident, e.g., in Figure 4 adapted from 
Amati et al. 2019, and in uptodate compilations of neasured 
$E_{iso}$ values of GRBs with known resdhif (e.g., Figure 8  
in Rossi et al. 2022). 

\begin{figure}[]
\centering 
\epsfig{file=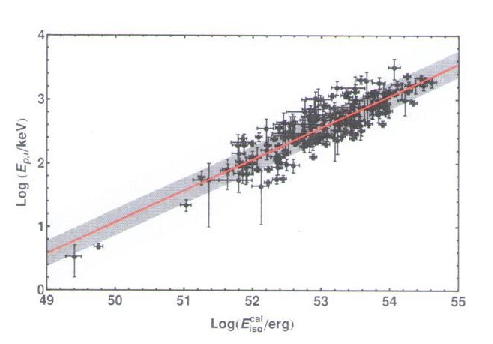,width=8.6cm,height=8.6cm}
\caption{The best fit Amati correlation (red line) between 
recalibrated values of $(1\!+\!z)E_p$ and  $E_{iso}$ of GRBs 
(black data points), within $1\sigma$ and $3\sigma$ limits 
(shaded region) adapted from Amati et al. 2019}
\end{figure}

\section{onclusions:}
In the CB model, GRBs and CRBs are produced by highly relativistic jets of 
plasmoids (cannonballs) ejected by fall back material in stripped envelope 
supernova explosions of massive stars. The knee, which has been discovered 
recently in the energy spectrum of high energy cosmic ray electrons + 
positrons, implies a maximum peak energy $(1\!+\!z)E_p \approx\!2.25$ MeV 
of GRB photons produced by ICS of glory photons near source. The Amati 
relation for such a peak photon energy, consistent with that predicted by 
the CB model, yields a maximum GRB isotropic equivalent energy 
$E_{iso}\!\approx\! 3.8\times 10^{54}$ erg. This value is consistent with 
the current highest value, $E_{iso}\!\approx \!3.7\times 10^{54}$ erg, 
(Atteia 2022) measured in GRB 220101A (Lesage et al. 2022) at redshift 
$z\!=\!4.618$ and with an earlier conclusion (Atteia 2017) that the 
distribution of the isotropic equivalent energy of GRBs has a strong 
cutoff above $1-3\times\! 10^{54}$ erg. This success provides further 
support to the validity of the unified Cannonball model of cosmic ray 
bursts and gamma ray bursts and in particular to the conclusion that the 
knees in the energy spectra of cosmic ray particles is proportional to 
their masses (Dar \& De R\'ujula 2008 and references therein) rather than 
to their rigidities, as widely believed.


\begin{references}

\reference {Abdollahi 2017}
Abdollahi, S. et al. (Fermi LAT Collab.) 2017, 
Phys. Rev. D {\bf 95}, 082007.

\reference {Adriani 2018}
Adriani, O. et al. (CALET Collab.) 2018, PRL {\bf 120}, 261102 
[arXiv:1806.0972].

\reference {Aguilar}
Aguilar, M.  et al. (AMS Collab.) 2014, PRL {\bf 113}, 221102.

\reference{Aharonian1}
Aharonian, F. et al. (H.E.S.S. Collab.) 2008, PRL {\bf 101}, 261104.

\reference {Aharonian2}
Aharonian, F.  et al. (H.E.S.S. Collab.) 2009, A\&A {\bf 508}, 561.

\reference{Amati1}
Amati, L.,  Frontera, F.,  Tavani, M., et al. 
2002, A\&A, 390, 81 [arXiv:astro-ph/0205230].

\reference{Amati2}
Amati, L., 2006, MNRAS, 372, 233 [arXiv:astro-ph/0601553].

\reference{Amati3}
Amati, L.,  Frontera, F.,  Guidorzi, C.  
2009, A\&A, 508, 173 [arXiv:0907.0384].

\reference{Amati4}
Amati, L. D'Agostino, R. Luongo, O. et al., 2019, 
MNRAS {\bf 486}, L46 (2019) [arXiv:1811.08934].

\reference {Ambrosi}
Ambrosi, G.  An, Q.  Asfandiyarov, R.  et al., 2017, Nature {\bf 552}, 63   
[arXiv:1711.10981].

\reference{Atteia1}
Atteia, J. L.  Heussaff, V.  Dezalay, J. P. et al., 2017, ApJ {\bf 837}, 
119  [arXiv:1702.02961].

\reference{Atteia2}
Atteia, J. L. 2022, GCN 31365.

\reference {Chang}
Chang, J. et al. (The DAMPE collaboration),
2017, Atropart. Phys. {\bf 95}, 6.

\reference{Costa}
Costa, E.  et al., 1997,  Nature {\bf 38}, 783  
[arXiv:astro-ph/9706065].

\reference{DD1}
Dado, S. Dar, A. De R`ujula, A. 2002,  A\&A {\bf 388}, 1079D  
[arXiv:astro-ph/0107367].

\reference{DD2}
Dado, S. Dar, A. 2015, ApJ {\bf 812}, 38  [arXiv:1502.01244].

\reference{DDD1}
Dado, S.  Dar, A. De R`ujula, A. 2022,  e print [arXiv:2204.04128].

\reference{Dar1}
Dar, A. 1998,  ApJ {\bf 500}, L93  [arXiv:astro-ph/9709231]. 

\reference{Dar2}
Dar, A. 1999,  A\&AS {\bf 138}, 505  [arXiv:astro-ph/9902017]. 

\reference{Dar3}
Dar, A.  Plaga, R. 1999, A\&A {\bf 349}, 259  
[arXiv:astro-ph/9902138]. 

\reference{Dar4}
Dar, A. De R`ujula, A. 2000,  e-print [arXiv:astro-ph/0012227].

\reference{DD01}
Dar, A. De R`ujula, A.  2004, Phys. Rep. {\bf 405}, 203  
[arXiv:astro-ph/0308248]. 

\reference{DD02}
Dar, A. De R\'ujula, A. 2008, Phys. Rept. {\bf 466}, 179  
[arXiv:hep-ph/0606199].

\reference{DeRujula}
De R\'ujula, A. 2019, PLB {\bf 790C}, 444  [arXiv:1802.06626].

\reference{Fishman}
Fishman, G. J. Meegan, C. A.  1995,  ARA\&A {\bf 33}, 415 .

\reference{Fu}
Fu, S. Y. Zhu, Z. P. Xu, D.   et al. 2022, GCN 31353.

\reference{Fynbo}
Fynbo, J. P. U. de Ugarte Postigo, A.  Xu, D.  et al. 2022, GCN 31359.


\reference{Klebesadel}
Klebesadel, R. W. Strong, I. B.  Olson, R. A.  1973, ApJ {\bf 182}, L85 .

\reference{Lessage}
Lesage, S. \& Meegan C. et al. (Fermi GRB Monitor Team) 2022, GCN 31360.


\reference{MacFadyen}
MacFadyen, A. Woosley, S. E. 1999,  ApJ {\bf 524}  262 
[arXiv:astro-ph/9810274].


\reference{Meegan}
Meegan, C. A. et al., 1992,  Nature  {\bf 355}, 143.


\reference{Rossi}
Rossi, A.  Frederiks, D. D. Kann, D. A. et al., 2022,  e-print 
[arXiv:2202.04544].

\reference{Shaviv}
Shaviv, N. J.  Dar, A.  1995,  ApJ, {\bf 447}, 863  
[arXiv:astro-ph/9407039].

\reference{van Paradijs}
van Paradijs, J.  et al., 1997, Nature, {\bf 386}, 686

\end{references}
\end{document}